\documentclass[a4paper,12pt]{article}
\usepackage[cp1251]{inputenc}
\usepackage{amsmath, amsthm, amsfonts, amssymb}
\usepackage[dvips]{graphicx}

\theoremstyle{plain}
\newtheorem{theorem}{Theorem}[section]

\newtheorem{lemma}{Lemma}[section]
\newtheorem{definition}{Definition}[section]
\newtheorem{proposition}{Proposition}[section]

\theoremstyle{remark}
\newtheorem{remark}{Remark}[section]

\numberwithin{equation}{section}

\begin{document}
\allowdisplaybreaks
\begin{center}
{\LARGE ON BOUNDARY -- VALUE PROBLEMS} \vskip10pt
{\LARGE FOR THE LAPLACIAN IN BOUNDED} \vskip10pt
{\LARGE AND IN UNBOUNDED DOMAINS WITH} \vskip10pt
{\LARGE PERFORATED BOUNDARIES}
\bigskip

\textbf{ Gregory A.Chechkin$^\natural$, \ Rustem
R.Gadyl'shin$^\flat$$^\sharp$ }

\end{center}
\bigskip
\centerline{$^\natural$ Department of Differential Equations}
\centerline{Faculty of Mechanics and Mathematics}
\centerline{Moscow State University}
\centerline{Moscow 119899, Russia}
\centerline{\tt chechkin{@}mech.math.msu.su}
\bigskip
\centerline{$^\flat$ Institute of Mathematics with Computing Center}
\centerline{Russian Academy of Sciences}
\centerline{ Ufa 450077, Russia}
\medskip
\centerline{$^\sharp$ Department of Mathematical Analysis}
\centerline{Faculty of Physics and Mathematics}
\centerline{Bashkir State Pedagogical University}
\centerline{ Ufa 450000, Russia}
\centerline{\tt gadylshin{@}bspu.ru}
\bigskip
\centerline {\bf CORRESPONDENCE ADDRESS:}
\centerline {Department of Differential Equations,}
\centerline {Faculty of Mechanics and Mathematics}
\centerline {Moscow State University, Moscow 119899, Russia.}
\centerline {\tt chechkin{@}mech.math.msu.su}
\centerline {Phone: +7-(095)-113-4535\ \ Fax: +7-(095)-939-2090}
\vskip40pt

\centerline{running head: {\bf ON BOUNDARY -- VALUE PROBLEMS ...}}

\def\thefootnote{ }
\footnote{The research of the first author was partially supported
by Russian Foundation for Basic Research (RFBR, project No
02-01-00868). The second author was partially supported by Russian
Foundation for Basic Research (RFBR, project No 02-01-00693) and
Scientific programm ``Universities of Russia"}

\newpage

\begin{center}
Abstract.
\end{center}

In the paper we consider boundary -- value problems with rapidly
alternating type of boundary conditions, including problems in
domains with perforated boundaries. We present the classification
of homogenized (limit) problems depending on the ratio of small
parameters, which characterize the diameter of parts of the
boundary with different types of boundary conditions or on the
ratio of small parameters, which characterize the diameter and the
distance between holes. Also we studied the analogue of the
Helmholtz resonator for domains with perforated boundary.

\def\thefootnote{ }
\footnote{2000. \textit{Mathematical Subject Classification.}
35B25, 35B27, 35J25} 

\setcounter{section}{-1}

\newpage
\section{Introduction}

We deal with boundary -- value problems with rapidly alternating
type of boundary conditions and problems in domains with
perforated boundaries. Problems of this kind attracted the
attention of mathematicians from the mid-1960-s (see, for
instance,~\cite{MKh} --~\cite{bib42}). Such problems appear in
physics and engineering sciences, when one studies, for example,
the scattering of acoustic waves on the small periodic obstacles,
the behavior of partially fastened membranes and many others. The
engineering applications of such problems could be also found in
construction of atomic power stations in space antennas etc. One
can study the problem of permeation of fuel through the walls of
plastic tank. In order to reduce permeation of fuel, the inner
boundary of the container is coated with thin barrier layer of
fluorine by a blow molding process. The resulting thin layer,
however, typically has flows: it leaves many small patches
uncovered. This model is described in more details in \cite{RN}
and \cite{F}.

In this work we consider boundary -- value problem in a 3D domain
for the Laplacian. We assume that the boundary of the domain
consists of two parts. One of them has purely periodic
microstructure. It could be rapidly alternating spots or
periodically situated holes. In the first case we have bounded
domain with micro inhomogeneous structure of the boundary and in
the second one we have two domains connected through the holes. In
the second case we study problems in both bounded domains or in
one bounded and one unbounded domains. We give a complete
classification of homogenized problems in their dependence on the
ratio of the small parameters, characterizing the frequency of the
periodical change of the boundary conditions in the first case or
on the ratio of the small parameters characterizing the diameter
and the distance between holes in the second case.

It should be noted that in the first case on the base of other
methods the convergence of solutions to such problems was proved
in~\cite{bib13} for more general situation. Also nonperiodic
boundary structure was considered in~\cite{bib2} and~\cite{bib23}.
On the other hand, the direct combination of the approaches
from~\cite{bib7} (see, also~\cite{bib6}) and~\cite{bib14}  (see,
also~\cite{bib15}) gives an opportunity to obtain the estimate for
the rate of convergence for solutions in the periodical situation
(see, also~\cite{bib8}). In this paper for the periodical boundary
microstructure we demonstrate the much shorter proof (than
in~\cite{bib13} and in ~\cite{bib7}  and~\cite{bib8}) of the
convergence theorem. We use the homogenization methods~\cite{BLP}
-- \cite{bib17} and the method of matching asymptotic expansions
\cite{bib27} -- \cite{bib16}. More precisely, we arrange the
approaches of~\cite{bib7} and~\cite{bib14}  in the most rational
form. This combination allows to get clear expressible formulae in
a short way.

Also we study problems in domains with perforated boundaries.
Similar problems were considered in \cite{MKh}, \cite{SP} --
\cite{A2}. In present paper we develop the rational approach
(mentioned above) for problems in domains with periodically
perforated boundaries. We find the ratio of the small parameters
characterizing the diameter and the distance between holes, that
implies in the limit the decomposition of the original problem to
two independent problems. In the case of unbounded external domain
we show that the decomposition involves the appearance of poles
(scattering frequencies) with small imaginary part of the
analytical continuation of solutions to the original problem. It
is wellknown that such poles for the Helmholtz resonator do exist
(see, for instance,~\cite{A} -- \cite{G2}). Namely these poles
induce the resonance in the Helmholtz resonator~\cite{A},
\cite{G4} -- \cite{G3}. Remind that the classical Helmholtz
resonator could be described by the boundary -- value problem for
the Helmholtz equation in an unbounded domain outside the surface
with a small aperture~\cite{bib78}, \cite{bib76}. The model
2-dimensional analogue of the Helmholtz resonator in periodically
perforated domain was considered in~\cite{bib29},\cite{bib31},
\cite{bib30}. The author discovered the resonances for this
analogue of the Helmholtz resonator. In this paper we consider
3-dimensional analogue of the Helmholtz resonator in
homogenization theory. We proved the existence of the scattering
frequencies with small imaginary part.

In the first section we introduce notation, describe the domains,
set the problems and formulate six basic theorems and one
auxiliary theorem which is of independent interest. The last
theorem is proved in the Section~\ref{s2}.

Problems with rapidly alternating boundary conditions is
considered in Sections~\ref{s3} and~\ref{s4}.

Sections~\ref{s5} -- \ref{s8} are devoted to investigations of
problems in domains with perforated boundaries. In
Sections~\ref{s5}, \ref{s6} we consider the case when connected
domains are bounded and in Sections~\ref{s7}, \ref{s8} we study
the case when one of the domains is unbounded, a 3D analogue of
the Helmholtz resonator in homogenization.


\section{Statements}\label{s1}


Let $\Omega$ be a bounded domain in ${\mathbb R}^3$ with a
$C^\infty$--boundary $\Gamma$. We suppose that $\Omega$ lies in
the half-space $x_3<0$,
$\Gamma_1=\mathrm{int}\left(\{x:\,x_3=0\}\cap \Gamma\right)$,
$\mathrm{mes}_2\Gamma_1\not=0$. Denote by $\omega$ a
two-dimensional bounded domain with a smooth boundary on the plane
$x_3=0$. We suppose that $0<\varepsilon, \, \delta<<1$ are small
parameters. Introduce the following notation:
$\omega_\varepsilon=\{x:\,x\varepsilon^{-1}\in\omega\}$,
$\Pi_\varepsilon=\{x:\,x=(2n,2m,0)+x',\
x'\in\omega_\varepsilon,\,n,m\in {\mathbb Z}\}$,
$\Pi_\varepsilon^\delta=\{x:\,\delta^{-1}x\in\Pi_\varepsilon\}$,
$\Gamma_2= \Gamma\backslash\overline{\Gamma_1}$,
$\Gamma_{\varepsilon,\delta}^D=\Gamma_1\cap\Pi_\varepsilon^\delta
$, and
$\Gamma_{\varepsilon,\delta}^S=\Gamma_1\backslash\overline
{\Gamma_{\varepsilon, \delta}^D}$ (see Fig. 1.).



\begin{figure}[htb]\label{fig.1}
\begin{center}
\includegraphics[height=8 true cm, width=10.1 true cm]{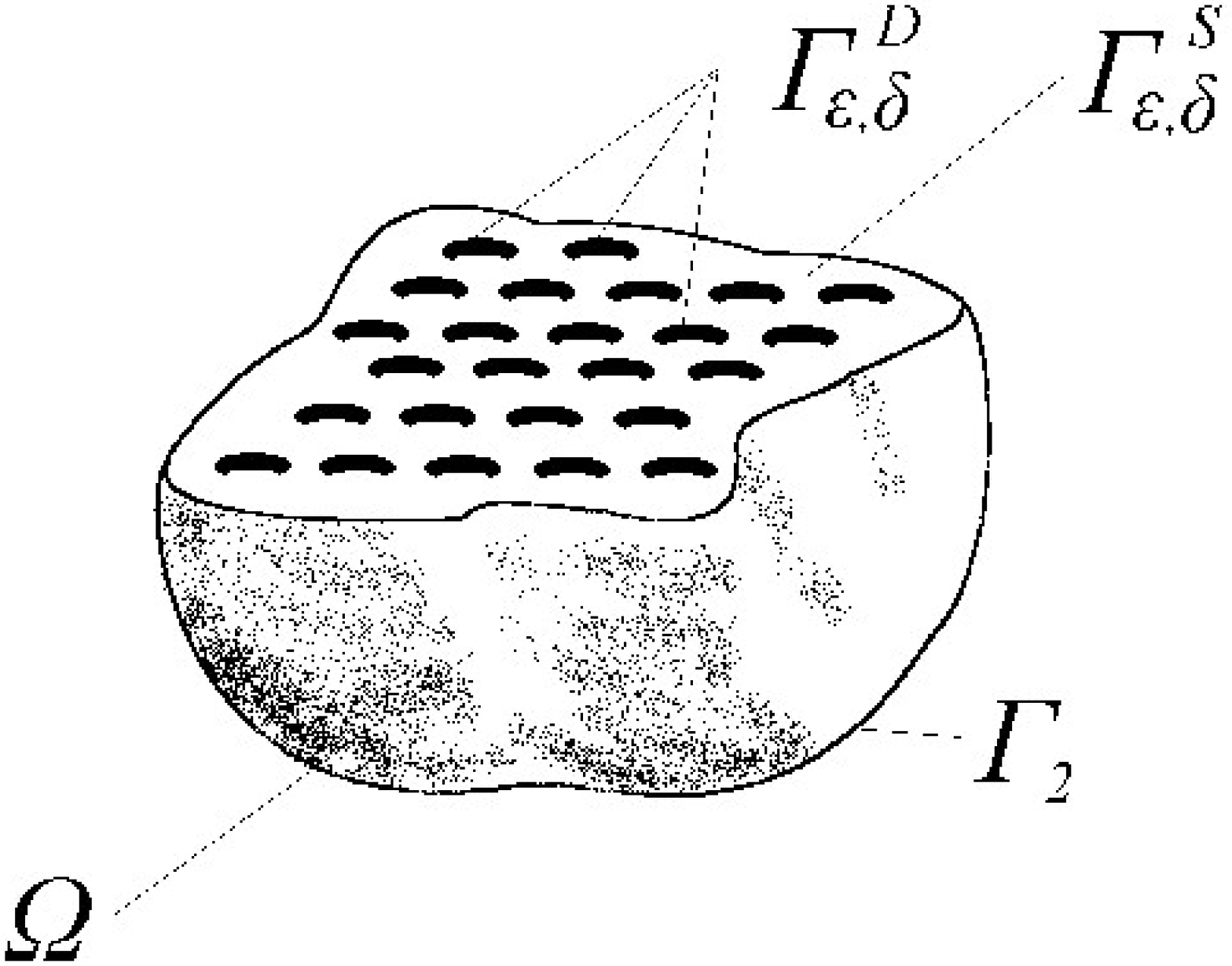}
\end{center}
\caption{}
\end{figure}

We consider the case when $\delta=\delta(\varepsilon)$ depends on
$\varepsilon$ and
\begin{equation*}
\lim_{\varepsilon\to0}\frac{\varepsilon}{\delta(\varepsilon)}=p,\qquad
p\in[0,\infty].
\end{equation*}

Our goal of the first part of  the work is to prove the following
two auxiliary statement:

\begin{theorem}\label{th1.1} Let $f\in L_2(\Omega)$, $q\ge0$,
$p=\infty$. Then the solution of the boundary value problem
\begin{equation}
\left\{
\begin{aligned}
&\Delta u_{\varepsilon,\delta}=f,\quad x\in\Omega, \cr
 & u_{\varepsilon,\delta}=0,\quad
 x\in\Gamma_{\varepsilon,\delta}^D\cup\Gamma_2, \cr & \dfrac{\partial
 u_{\varepsilon,\delta}}{\partial x_3}+qu_{\varepsilon, \delta}=0, \quad
x\in\Gamma_{\varepsilon,\delta}^S
\end{aligned}
\right. \label{1.1}
\end{equation}
converges to the solution of the boundary -- value problem
\begin{align}
\left\{
\begin{aligned}
 \Delta & u_0=f,\quad x\in\Omega, \cr & u_0=0,\quad x\in\Gamma
\end{aligned}
\right. \label{1.3}
\end{align}
in $H^1(\Omega)$.
\end{theorem}

\begin{theorem}\label{th1.2} Let
$f\in L_2(\Omega)$, $q\ge0$, $p<\infty$, $Q=q+c_\omega p$. Then
the solution of {\rm (\ref{1.1})} converges to the solution of the
boundary -- value problem
\begin{align}
\left\{
\begin{aligned}
 \Delta & u_0=f,\quad x\in\Omega,\cr & u_0=0,\quad x\in\Gamma_2,
\cr & \dfrac{\partial  u_0}{\partial x_3}+Qu_0=0,\quad
x\in\Gamma_1.
\end{aligned}
\right.\label{1.4}
\end{align}
in $H^1(\Omega)$, if $p=0$, and weakly in $H^1(\Omega)$ and
strongly in $L_2(\Omega)$, if $p>0$.
\end{theorem}

Hereafter,  $c_\omega$ is the capacity of the plate $\omega$
(\cite{bib19},~\cite{bib25}). This wellknown constant is positive
and, for instance, if $\omega$ is the unit disk, then
$c_\omega=2\pi^{-1}$ (see,~\cite{bib19}).

Later on we consider problems in domains with perforated boundary.
In Sections~5 and 6 we study problems in bounded domains.

Let $\widetilde\Omega$ be a bounded domain in ${\mathbb R}^3$ with
a $C^\infty$--boundary $\widetilde\Gamma$ such that
$\overline{\Omega}\subset\widetilde\Omega$ (see Fig. 2.).
\begin{figure}[htb]\label{fig.2}
\begin{center}
\includegraphics[height=10 true cm, width=11.56 true cm]{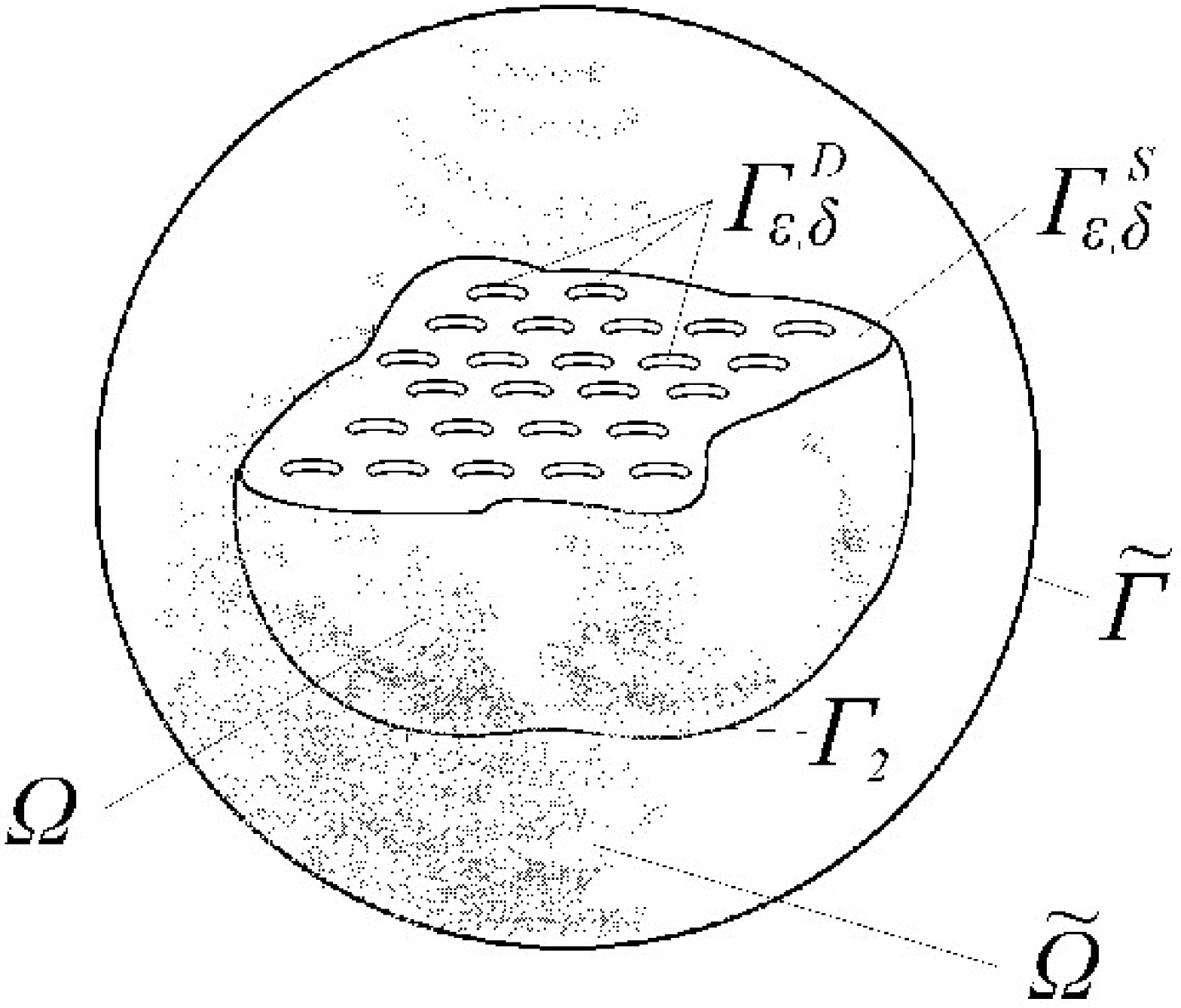}
\end{center}
\caption{}
\end{figure}

\begin{theorem}\label{th1.4} Let  $p=\infty$,
$F\in L_2(\widetilde\Omega)$, $f$ and $\widetilde{f}$ be the
restrictions of $F$ in $\Omega$ and in
$\widetilde\Omega\backslash\overline{\Omega}$, respectively. Then
the solution of the boundary -- value problem
\begin{align}
\left\{
\begin{aligned}
&\Delta u_{\varepsilon,\delta}=F,\quad
x\in\widetilde\Omega\backslash\overline{\Gamma_{\varepsilon,\delta}^D\cup\Gamma_2},
\cr
 & u_{\varepsilon,\delta}=0,\quad x\in\Gamma_{\varepsilon,\delta}^D\cup\Gamma_2\cup\widetilde
 \Gamma,
\end{aligned}\right.
\label{5.1}
\end{align}
converges strongly in $H^1(\widetilde \Omega)$ as
$\varepsilon\to0$ to the function
\begin{align*}
u(x)= \left\{
\begin{aligned}
&u_0(x),\quad x\in\Omega, \cr &\widetilde u_0(x),\quad
x\in\widetilde\Omega\backslash\overline \Omega,
\end{aligned}\right.
\end{align*}
where $u_0(x)$ is a solution of Problem {\rm (\ref{1.3})} and
$\widetilde u_0(x)$ satisfies the boundary -- value problem
\begin{align}
 \left\{
\begin{aligned}
& \Delta \widetilde{u}_0=\widetilde{f},\quad
x\in\widetilde\Omega\backslash\overline
 {\Omega}, \cr &
 \widetilde{u}_0=0,\quad x\in\Gamma\cup\widetilde\Gamma.
\end{aligned}\right.
\label{5.3}
\end{align}
\end{theorem}

\begin{theorem}\label{th1.5} Let  $p=0$,
$F\in L_2(\widetilde\Omega)$, $f$ and $\widetilde{f}$ be the
restrictions of $F$ in $\Omega$ and in
$\widetilde\Omega\backslash\overline{\Omega}$, respectively. Then
the solution of the boundary -- value problem
\begin{align}
 \left\{
\begin{aligned}
& \Delta u_{\varepsilon,\delta}=F,\quad
x\in\widetilde\Omega\backslash\overline{\left(\Gamma_2\cup\Gamma_{\varepsilon,\delta}^S
\right)}, \cr & u_{\varepsilon,\delta}=0,\quad
x\in\Gamma_2\cup\widetilde\Gamma, \cr & \dfrac{\partial
u_{\varepsilon,\delta}}{\partial x_3}=0, \quad
x\in\Gamma_{\varepsilon,\delta}^S,
\end{aligned}\right.\label{6.1}
\end{align}
converges strongly in $H^1(\widetilde \Omega\backslash \overline
\Gamma_1)$ as $\varepsilon\to0$ to the function
\begin{align*}
u(x)= \left\{
\begin{aligned}
&u_0(x),\quad x\in\Omega, \cr &\widetilde u_0(x),\quad
x\in\widetilde\Omega\backslash\overline \Omega,
\end{aligned}\right.
\end{align*}
where $u_0(x)$ is a solution of the boundary -- value problem
\begin{align} & \left\{
\begin{aligned}
& \Delta u_0=f,\quad x\in\Omega,\cr &
 u_0=0,\quad x\in\Gamma_2, \cr &
 \dfrac{\partial u_0}{\partial x_3}=0,\quad x\in\Gamma_1,
\end{aligned}\right.
\label{6.2}
\end{align}
and $\widetilde u_0(x)$ is a solution of the boundary -- value
problem
\begin{align}
& \left\{
\begin{aligned}
& \Delta \widetilde{u}_0=\widetilde{f},\quad
x\in\widetilde\Omega\backslash\overline{\Omega},\cr &
 \widetilde{u}_0=0,\quad x\in\Gamma_2\cup\widetilde \Gamma, \cr &
 \dfrac{\partial \widetilde{u}_0}{\partial x_3}=0,\quad x\in\Gamma_1.
\end{aligned}\right.
\label{6.3}
\end{align}
\end{theorem}

Denote by $\sigma$ the square $(-1,1)\times(-1,1)$ in the plane
$x_3=0$ and let $\Sigma=\sigma\times(0,-\infty)$. Denote by
$C^\infty(\overline\Sigma,\omega_\varepsilon)$ the set of
$C^\infty$--functions vanishing in a neighborhood of
$\omega_\varepsilon$ and having the finite Dirichlet integral. We
define the space $\widetilde H^1(\Sigma; \omega_\varepsilon)$ as
the closure of $C^\infty(\overline\Sigma)$ by the norm
\begin{equation*}
\|v\|_1=\left(\int\limits_\Sigma|\nabla
v|^2\,dx+\int\limits_\sigma v^2\,ds\right)^{1/2}.
\end{equation*}

The proofs of Theorems~\ref{th1.1} and \ref{th1.4} are based on
the following statement.

\begin{theorem}\label{th1.3} Asymptotics of
\begin{equation}
 \lambda_\varepsilon=\inf_{v\in\widetilde
H^1(\Sigma;\omega_\varepsilon)\backslash
\{0\}}\frac{\int\limits_\Sigma|\nabla v|^2\,dx}{\int\limits_\sigma
v^2\,ds}\label{1.7}
\end{equation}
reads as follows:
\begin{equation*}
\lambda_\varepsilon=\varepsilon\frac{\pi
c_\omega}{2}+o(\varepsilon).
\end{equation*}
\end{theorem}

The last part of the paper is devoted to the investigation of
problems  in unbounded domains. Assume that $F$ is a function from
$L_2({\mathbb R}^3)$ with bounded support. We consider the
following boundary -- value problems:
\begin{align}
\left\{
\begin{aligned}
&\left(\Delta+k^{2}\right) u_{\varepsilon,\delta}=F,\quad
x\in\mathbb{R}^{3}\backslash\overline{\Gamma_{\varepsilon,\delta}^D\cup\Gamma_2},
\cr
 & u_{\varepsilon,\delta}=0,\quad
 x\in\Gamma_{\varepsilon,\delta}^D\cup\Gamma_2,
\end{aligned}\right.
\label{1.10}
\end{align}
\begin{align}
\left\{
\begin{aligned}
&\left(\Delta+k^{2}\right) u_{\varepsilon,\delta}=F,\quad
x\in\mathbb{R}^{3}\backslash\overline{\left(\Gamma_2\cup\Gamma_{\varepsilon,\delta}^S
\right)}, \cr
 & u_{\varepsilon,\delta}=0,\quad
 x\in\Gamma_2,\qquad \dfrac{\partial
u_{\varepsilon,\delta}}{\partial x_3}=0, \quad
x\in\Gamma_{\varepsilon,\delta}^S,
\end{aligned}\right.
\label{1.11}
\end{align}
with the radiation condition
\begin{equation}\label{rad1}
 u_{\varepsilon,\delta}=O(r^{-1}),\qquad \frac{\partial
u_{\varepsilon,\delta}}{\partial r} -
iku_{\varepsilon,\delta}=o(r^{-1}),\qquad r\to\infty.
\end{equation}
for $k$ such that ${\rm Im}\,k\geq 0$. Here and throughout
$r=|x|$.

\begin{theorem}\label{th1.6} Let  $p=\infty$. Suppose also that $f$ and $\widetilde{f}$ are the
restrictions of $F$ in $\Omega$ and in
$\mathbb{R}^{3}\backslash\overline{\Omega}$, respectively. Then
the solution to Problem {\rm (\ref{1.10})},
{\rm (\ref{rad1})} converges strongly in $H^1_{\rm loc}(\mathbb
R^3)$ as $\varepsilon\to0$ to the function
\begin{align*}
u(x)= \left\{
\begin{aligned}
&u_0(x),\quad x\in\Omega, \cr &\widetilde u_0(x),\quad
x\in\widetilde\Omega\backslash\overline \Omega,
\end{aligned}\right.
\end{align*}
where $u_0(x)$ is a solution of the boundary -- value problem
\begin{align} \left\{
\begin{aligned}
-\Delta & u_0=k^2u_0-f,\quad x\in\Omega, \cr & u_0=0,\quad
x\in\Gamma,
\end{aligned}
\right. \label{1.33}
\end{align}
and $\widetilde u_0(x)$ is a solution of the boundary -- value
problem
\begin{align}
\left\{
\begin{aligned}
&\left(\Delta+k^{2}\right) \widetilde{u}_0=\widetilde{f},\quad
x\in\mathbb{R}^{3}\backslash\overline{\Omega}, \cr
 & \widetilde{u}_0=0,\quad
 x\in\Gamma,
\end{aligned}\right.
\label{1.12}
\end{align}
with the radiation condition
\begin{equation}\label{rad2}
 \widetilde{u}_0=O(r^{-1}),\qquad \frac{\partial \widetilde{u}_0}{\partial r}
- ik\widetilde{u}_0=o(r^{-1}),\qquad r\to\infty.
\end{equation}

Here it is assumed that $k^2$ is not an eigenvalue of Problem {\rm
(\ref{1.33})}.

If $k^2=k^2_{0}$ is an eigenvalue to Problem
{\rm (\ref{1.33})}, then there is a pole $\tau_{\varepsilon}$ of
the analytic continuation of the solution of  {\rm (\ref{1.10}),
(\ref{rad1})} in the half plane ${\rm Im}\, k<0,$ converging to
$k_{0}$ as $\varepsilon\to0$.
\end{theorem}

\begin{theorem}\label{th1.7} Let  $p=0$,  $f$ and $\widetilde{f}$ be the
restrictions of $F$ in $\Omega$ and in
$\mathbb{R}^{3}\backslash\overline{\Omega}$, respectively. Then
the solution to Problem (\ref{1.11}),
(\ref{rad1}) converges strongly in $H^1_{\rm loc}(\mathbb
R^3\backslash\overline \Gamma_1)$ as $\varepsilon\to0$ to the
function
\begin{align*}
u(x)= \left\{
\begin{aligned}
&u_0(x),\quad x\in\Omega, \cr &\widetilde u_0(x),\quad
x\in\widetilde\Omega\backslash\overline \Omega,
\end{aligned}\right.
\end{align*}
where $u_0(x)$ is a solution of the boundary -- value problem
\begin{align} & \left\{
\begin{aligned}
& -\Delta u_0=k^2u_0-f,\quad x\in\Omega,\cr &
 u_0=0,\quad x\in\Gamma_2, \cr &
 \dfrac{\partial u_0}{\partial x_3}=0,\quad x\in\Gamma_1
\end{aligned}\right.
\label{6.22}
\end{align}
and $\widetilde u_0(x)$ is a solution of the boundary -- value
problem
\begin{align}
\left\{
\begin{aligned}
&\left(\Delta+k^{2}\right) \widetilde{u}_0=\widetilde{f},\quad
x\in\mathbb{R}^{3}\backslash\overline{\Omega}, \cr
 & \widetilde{u}_0=0,\quad
 x\in\Gamma_{2},\qquad \frac{\partial \widetilde{u}_0}{\partial x_{3}}=0,\quad
 x\in\Gamma_{1}
\end{aligned}\right.
\label{1.13}
\end{align}
with the radiation condition {\rm (\ref{rad2})}.

Here it is assumed that $k^2$ is not an eigenvalue of Problem {\rm
(\ref{6.22})}.

If $k^2=k^2_{0}$ is an eigenvalue of the boundary -- value problem
{\rm (\ref{6.22})}, then there is a pole $\tau_{\varepsilon}$ of
the analytic continuation of the solution of  {\rm (\ref{1.11}),
(\ref{rad1})} in the half plane ${\rm Im}\, k<0,$ converging to
$k_{0}$ as $\varepsilon\to0$.
\end{theorem}

The notion of an analytic continuation is classical and we shall
give all the necessary definitions in section \ref{s7}.

\section{ Proof of Theorem~\ref{th1.3}}\label{s2}


Suppose that $G\in C^\infty(\overline\Sigma\backslash\{0\})$ is a bounded
$\sigma$--periodic solution of the boundary -- value problem
\begin{equation}
 \Delta G=0,\quad x\in\Sigma,\qquad \dfrac{ \partial G}{ \partial x_3}
=-\frac{1}{4}, \quad x\in\sigma\backslash\{0\},\label{2.1}
\end{equation}
with the asymptotics $G(x)\sim(2\pi r)^{-1}$ as $r=|x|\to0$. The
existence theorem of this Green function is wellknown. In more
detail the asymptotics of $G$ reads as follows (see, for instance,
~\cite{bib1} ):
\begin{equation}
\begin{aligned}
G(x) & =C_\Sigma+O\left(\operatorname{exp}\left\{\frac{\pi
x_3}{2}\right\}\right), \qquad x_3\to-\infty,
\\
G(x) & =\frac{1}{2\pi r}-\frac{1}{4}x_3+O\left(r^2\right),\qquad
r\to0,
\end{aligned}\label{2.2}
\end{equation}
where $C_\Sigma$ is a fixed constant. It is easy to show that the
right--hand side of the asymptotics (\ref{2.2}) are a
differentiable function. In view of the evenness the function $G$
satisfies the homogeneous Neumann boundary condition in
$\partial\Sigma\backslash\overline\sigma$.

It is known (see, for instance~\cite{bib12} ) that there exists a
harmonic in ${\mathbb R}^3_-=\{x:\,x_3<0\}$ function $X_0\in
H^1_{loc}({\mathbb R}^3_-)\cap C^\infty(\overline{{\mathbb
R}^3_-}\backslash
\partial\omega)$ which vanishes at infinity and satisfies the
boundary conditions $X_0=1$ in $\omega$ and $ \partial X_0 /
\partial x_3=0$ on $\gamma=\{x:\,x_3=0,\,\,(x_1,x_2)\notin\overline
\omega\}$. In addition the function $X_0$ has the differentiable
asymptotics
\begin{equation}
 X_0(x)=c_\omega r^{-1}+\sum_{i=1}^2 c_i\frac{
\partial}{\partial x_i}r^{-1} +O(r^{-3})\qquad\text{as
$r\to\infty$}. \label{2.3}
\end{equation}

It should be noted, that the capacity of $\omega$ is defined
namely as the coefficient of $r^{-1}$ in (\ref{2.3}).

Denote by $\chi(t)$ a smooth cut-off function equals to one for
$t<1/3$ and equals to zero for $t>2/3$ and suppose that
\begin{equation*}
\widetilde W^\varepsilon(x,\beta)=1- \varepsilon 2\pi
c_\omega\left(1-\chi(r\varepsilon^{-\beta})\right) G(x)-
 \chi(r\varepsilon^{-\beta})
X_0(x\varepsilon^{-1}),
\end{equation*}
where $\beta>0$. Taking into account (\ref{2.1})--(\ref{2.3}), one
can obtain the following statement.

\begin{lemma}\label{lm2.1} The function
$\widetilde W^\varepsilon\in
\widetilde H^1(\Sigma;\omega_\varepsilon)\cap C^\infty(\overline\Sigma\backslash
 \partial\omega_\varepsilon)$
satisfies the estimates
\begin{equation*}
\|\widetilde W^\varepsilon-1\|_{L_2(\sigma)}=o(1),
\quad\varepsilon\to0,\qquad |\nabla\widetilde W^\varepsilon|=
O\left(\operatorname{exp}\left\{\frac{\pi x_3}{2}\right\}\right),
\quad x_3\to-\infty
\end{equation*}
and it is a solution to the boundary -- value problem
\begin{equation*}
\left\{\begin{aligned} &\Delta\widetilde W^\varepsilon=\widetilde
F_\varepsilon,\quad x\in\Sigma,\cr & \dfrac{ \partial\widetilde
W^\varepsilon}{
\partial x_3}=
 \varepsilon\dfrac{\pi}{2} c_\omega\widetilde W^\varepsilon+
 \widetilde h_\varepsilon,\quad x\in\sigma\backslash \overline{\omega_
 \varepsilon},\cr &
\widetilde W^\varepsilon=0,\quad x\in\omega_\varepsilon, \cr &
\dfrac{ \partial\widetilde W^\varepsilon}{ \partial\nu}=0,\quad
x\in
\partial\Sigma \backslash\overline\sigma,
\end{aligned}\right.
\end{equation*}
where $\mathrm{supp}\widetilde F_\varepsilon$ lies above the plane
$x_3=-\frac{2}{3}\varepsilon^\beta$.

If $\beta<2/5$, then
\begin{equation*}
\|\widetilde F_\varepsilon \|_{L_2(\Sigma)}+\|\widetilde
h_\varepsilon\|_{L_2( \sigma)}= o(\varepsilon),
\qquad\varepsilon\to0.
\end{equation*}
\end{lemma}

By a standard way (see, for instance,~\cite{bib7} ) it is easy to
show that infimum in (\ref{1.7}) is attained at some harmonic
function $W^\varepsilon$ and
\begin{equation}
 \int\limits_\Sigma(\nabla
W^\varepsilon,\nabla v)\,dx=\lambda_\varepsilon\int\limits_\sigma
W^ \varepsilon v\,ds\label{2.4}
\end{equation}
for any $v\in\widetilde H^1(\Sigma;\omega_\varepsilon)$, where
$\lambda_\varepsilon$ is defined in (\ref{1.7}).

Due to Lemma~\ref{lm2.1} we have
\begin{equation}
\int\limits_\Sigma\widetilde F_\varepsilon
W^\varepsilon\,dx-\int\limits_\sigma \widetilde h_\varepsilon
W^\varepsilon\,ds=-\int\limits_\Sigma(\nabla\widetilde
W^\varepsilon, \nabla W^\varepsilon)\,dx+\varepsilon\frac{\pi
c_\omega}{2} \int\limits_\sigma\widetilde W^\varepsilon
W^\varepsilon\,ds.\label{2.5}
\end{equation}
From (\ref{2.4}) and (\ref{2.5}) we deduce the expression
\begin{equation}
 \int\limits_\Sigma\widetilde F_\varepsilon
W^\varepsilon\,dx-\int\limits_\sigma \widetilde h_\varepsilon
W^\varepsilon\,ds=\left(\varepsilon\frac{\pi
c_\omega}{2}-\lambda_\varepsilon\right)
\int\limits_\sigma\widetilde W^\varepsilon W^
\varepsilon\,ds.\label{2.6}
\end{equation}
Denote by $\tau$ the maximum $\max\limits_{x'\in\omega}\{|x'|\}$.
Obviously, the function
\begin{equation*}
w^\varepsilon(x)=1-\chi(3\tau\varepsilon^{-1}r)
\end{equation*}
belongs to $\in\widetilde H^1(\Sigma;\omega_ \varepsilon)$ and $
\|\nabla
w^\varepsilon\|_{L_2(\Sigma)}+\|w^\varepsilon-1\|_{L_2(\sigma)}
\to0 $ as $\varepsilon\to0$. Hence,
\begin{equation}
 \lambda_\varepsilon\to0, \qquad
\varepsilon\to0.\label{2.7}
\end{equation}
We suppose that $W^\varepsilon$ is normalized, i.e.
$\int\limits_\sigma (W^ \varepsilon)^2\,ds=1$. It should be noted
that (\ref{2.7}) and (\ref{1.7}) lead to $ \|\nabla
W^\varepsilon\|_{L_2(\Sigma)}\to0.$  Hence, for any fixed $R<0$,
there exists a representation
$W^\varepsilon(x)=w^\varepsilon_R(x)+c_R(\varepsilon)$, where
$\|w^ \varepsilon_R \|_{H^1(\Sigma_R)}\to0$ as $\varepsilon\to0$
and $\Sigma_R=\Sigma\cup\{x:\,x_3>R\}$. Therefore,
$\|w^\varepsilon_R\|_{L_2(\sigma)}\to0$ as $\varepsilon\to0$ and
then $c_R(\varepsilon)\to1$. The latter convergences imply that
\begin{equation}
\|W^\varepsilon\|_{L_2(\Sigma_R)}\le C_R,\qquad
\|W^\varepsilon-1\|_{L_2(\sigma)}\to0,\qquad
\varepsilon\to0,\label{2.8}
\end{equation}
where the constant $C_R$ is independent of $\varepsilon$. Now
(\ref{2.6}), (\ref{2.8}) and Lemma~\ref{lm2.1} (with $\beta<2/5$)
prove Theorem~\ref{th1.3}.

\section{Proof of Theorem~\ref{th1.1}}\label{s3}

We consider Problem (\ref{1.1})--(\ref{6.3}) in a weak sense (see,
for instance,\cite{bib18}).

\begin{definition}
Denote by $C^\infty(Q;S)$ the set of functions from
$C^\infty(\overline{Q})$ vanishing in a neighborhood of
$S\subset\overline{Q}$.

Define the space ${H^1}(Q;S)$ as the closure of the set of
functions from $C^\infty(Q;S)$ by the norm of the Sobolev space
$H^1(Q)$.
\end{definition}

\begin{definition}
A function
$u_{\varepsilon,\delta}\in{H^1}(\Omega;\Gamma_{\varepsilon,
\delta}^D\cup\Gamma_2)$ is called a solution of (\ref{1.1}), if
there holds the integral identity
\begin{equation}
\int\limits_\Omega\nabla u_{\varepsilon,\delta}\nabla
v\,dx+q\int\limits_{\Gamma_{\varepsilon,\delta}^S}u_{\varepsilon,\delta}v\,ds=
-\int\limits_\Omega fv\,dx\label{1.2}
\end{equation}
for any
$v\in{H^1}(\Omega;\Gamma_{\varepsilon,\delta}^D\cup\Gamma_2)$.
\end{definition}

In a similar way we define solutions of the boundary -- value problem
(\ref{1.3}).

\begin{definition}
The function $u_0\in{H^1}(\Omega;\Gamma)$ is a solution of
(\ref{1.3}), if it satisfies the identity
\begin{equation}
 \int\limits_\Omega\nabla u_0\nabla v\,dx=
-\int\limits_\Omega fv\,dx,\label{1.5}
\end{equation}
for any $v\in{H^1}(\Omega;\Gamma)$.
\end{definition}

Since the usual $H^1$-norm and the norm $
\|u\|'_{H^1(\Omega)}=\|\nabla u\|_{L_2(\Omega)}$ are equivalent in
${H^1}(\Omega;\Gamma_2)$ and ${H^1}(\Omega;\Gamma_{\varepsilon,
\delta}^D\cup\Gamma_2)\subset {H^1}(\Omega; \Gamma_2)$, we deduce
from (\ref{1.2}) the uniform estimate
\begin{equation}
\|u_{\varepsilon,\delta}\|_{H^1(\Omega)}\le
C\|f\|_{L_2(\Omega)}.\label{3.1}
\end{equation}

\begin{lemma}\label{lm3.1} For any
$v\in{H^1}(\Omega;\Gamma_{\varepsilon,\delta}^D\cup\Gamma_2)$ the
following estimate holds:
\begin{equation}
\|v\|_{L_2(\Gamma)}\le
C\left(\frac{\delta}{\varepsilon}\right)^{1/2}\|v\|_{H^1(\Omega)}.\label{3.2}
\end{equation}
\end{lemma}

\noindent\textbf{Proof.} Consider the extension of $v$ by zero
into ${\mathbb R}_-^3\backslash\Omega$. Obviously, this extension
belongs to $H^1({\mathbb R}^3_-)$. Assume that
$\sigma_\delta^{(m,n)}=\{x:\,x=\delta(2m,2n,0)+x',\,x'\delta^{-1}\in\sigma
\}$,
$\Sigma_\delta^{(m,n)}=\sigma_\delta^{(m,n)}\times(-\infty,0)$.
Denote by $\Sigma_\delta^ \partial$ the union of
$\Sigma_\delta^{(m,n)}$ such that $\sigma_\delta^{(m,n)}\cap
\partial\Gamma_1\not=\emptyset$, by $\Sigma_\delta^{in}$ the union
of $\Sigma_\delta^{(m,n)}$ such that $\sigma_\delta^{(m,n)}\subset
\Gamma_1$, and by $\Sigma_\delta^{ex}$ the union of
$\Sigma_\delta^{(m,n)}$ such that
$\sigma_\delta^{(m,n)}\cap\Gamma_1=\emptyset$. Then
\begin{equation}
\|v\|_{L_2(\sigma_\delta^{(m,n)})}\le
C\left(\frac{\delta}{\varepsilon}\right)^{1/2}\|\nabla
v\|_{L_2(\Sigma_\delta^ {(m,n)})},\qquad
\Sigma_\delta^{(m,n)}\subset\Sigma_\delta^{ex} \label{3.3}
\end{equation}
and, due to (\ref{1.7}) and Theorem~\ref{th1.3} the following
estimate takes place:
\begin{equation}
\|v\|_{L_2(\sigma_\delta^{(m,n)})}\le
C\left(\frac{\delta}{\varepsilon}\right)^{1/2}\|\nabla
v\|_{L_2(\Sigma_\delta^ {(m,n)})},\qquad
\Sigma_\delta^{(m,n)}\subset\Sigma_\delta^{in}. \label{3.4}
\end{equation}

Now, suppose that $\Sigma_\delta^{(m,n)}\subset\Sigma_\delta^
\partial$,
$\omega_\varepsilon^\delta=\{x:\,x=\delta(2m,2n,0)+x',\,x'\delta^{-1}\in
\omega\}$, and $\widetilde\omega_\varepsilon^\delta$ are the
subset of $\sigma_\delta^{(m,n)}$, where the function $v$ equals
to zero. Since,
$\omega_\varepsilon^\delta\subset\widetilde\omega_\varepsilon^\delta$,
then (\ref{1.7}) and Theorem~\ref{th1.3} imply that
\begin{equation}
\|v\|_{L_2(\sigma_\delta^{(m,n)})}\le
C\left(\frac{\delta}{\varepsilon}\right)^{1/2}\|\nabla
v\|_{L_2(\Sigma_\delta^ {(m,n)})},\qquad
\Sigma_\delta^{(m,n)}\subset\Sigma_\delta^ \partial. \label{3.5}
\end{equation}

Estimates (\ref{3.3})--(\ref{3.5}) give estimate (\ref{3.2}).
Lemma is proved.

Let $p=\infty$, suppose also that $v$ is a function from
$C^\infty_0(\Omega)$ and $\{\varepsilon_n\}$ is a sequence which
tends to zero as $n\to\infty$. Due to the embedding theorems, weak
compactness of the bounded set in $H^1(\Omega)$ and estimates
(\ref{3.1}), (\ref{3.2}) there exists a subsequence of this
sequence such that $u_{\varepsilon,\delta}\to u_0\in {H^1}
(\Omega;\Gamma)$ at this subsequence weakly in $H^1(\Omega)$ and
strongly in $L_2(\Omega)$. Passing to the limit in (\ref{1.2}) on
this subsequence we obtain (\ref{1.5}). Hence, $u_0$ is the
solution of (\ref{1.3}). On the other hand, due to the
arbitrariness of choosing of the sequence of $\{\varepsilon_n\}$,
we have that $u_{\varepsilon,\delta}\to u_0$ as $\varepsilon\to0$
(weakly in $H^1(\Omega)$ and strongly in $L_2(\Omega)$). Taking in
account the latter convergences and coming to the limit in
(\ref{1.2}) with $v=u_{\varepsilon,\delta}$, we obtain convergence
\begin{equation*}
\|u_{\varepsilon,\delta}\|_{H^1(\Omega)}\to \|u_0\|_{H^1(\Omega)}.
\end{equation*}
Hence, $u_{\varepsilon,\delta}\to u_0$ as $\varepsilon\to0$
strongly in $H^1(\Omega)$. Theorem is proved.

\section{Proof of Theorem~\ref{th1.2}}\label{s4}


Let us introduce the notation
\begin{equation*}
W_\varepsilon(x;\beta)= \widetilde W^\varepsilon(x;\beta)-
\varepsilon^2\,2\pi\left(1-\chi(r\varepsilon^{-\beta})\right)\sum_{i=1}^2
c_i\frac{ \partial}{\partial x_i}G(x) +\varepsilon \frac{\pi
c_\omega}{2}\chi(r\varepsilon^{-\beta})x_3.
\end{equation*}

After this correction of the function $\widetilde
W^\varepsilon(x;\beta)$ we obtain the following statement.

\begin{lemma}\label{lm4.1} The $\sigma$-periodic function
$W_\varepsilon(x;\beta)\in H^1_{loc}(\Sigma)
\cap C^\infty(\overline\Sigma\backslash \partial\omega_\varepsilon)$
satisfies the estimates
\begin{align*}
& \|W_\varepsilon-1+\varepsilon2\pi c_\omega
C_\Sigma\|_{L_2(\Sigma)}=o(1), \quad\varepsilon\to0,
\\
& |\nabla W_\varepsilon|+|W_\varepsilon-1+\varepsilon2\pi c_\omega
C_\Sigma|= O\left(\operatorname{exp}\left\{\frac{\pi
x_3}{2}\right\}\right), \quad x_3\to-\infty,\,\,\varepsilon\to0
\end{align*}
and $W_\varepsilon$ is a solution to the boundary -- value problem
\begin{equation*}
\left\{
\begin{aligned}
\Delta W_\varepsilon & =F_\varepsilon,\quad x\in\Sigma,\\
\displaystyle\frac{\partial W_\varepsilon}{\partial x_3} &
=
\varepsilon{\displaystyle\frac{\pi}{2}}c_\omega W_\varepsilon
+h_\varepsilon,\quad x\in\sigma\backslash
\overline{\omega_\varepsilon},\\ W_\varepsilon & =0,\quad
x\in\omega_\varepsilon,
\end{aligned}\right.
\end{equation*}
where $\mathrm{supp}F_\varepsilon$ lies above the plane
$x_3=-\frac{2}{3} \varepsilon^\beta$.

If $\frac{1}{5}<\beta<\frac{3}{7}$, then
\begin{equation*}
\|F_\varepsilon \|_{L_2(\Sigma)}= o(\varepsilon^{3/2}),\qquad
\|h_\varepsilon\|_{L_2(\sigma)}=o(\varepsilon).
\end{equation*}
\end{lemma}

\begin{remark}\label{rm4.1} We use our correction only to improve
estimates of the right--hand side of the equation. This estimate
will be used in the proof of Theorem~\ref{th1.2}.
\end{remark}

Denote by $W_\varepsilon^\delta(x)$ the expression
\begin{equation*}
1+\chi(-x_3\delta^{-1/2})\left(\frac{1}{1-\varepsilon2 \pi
c_\omega C_\Sigma} W_\varepsilon\left(\frac{x}{\delta};\frac{1}{4}
\right)-1\right).
\end{equation*}
Lemma~\ref{lm4.1} implies the following result, which is a key to
prove Theorem~\ref{th1.2}.

\begin{lemma}\label{lm4.2} If $\varepsilon\to0$ and $\delta\to0$, then
\begin{align*}
& \|\Delta
W_\varepsilon^\delta\|_{L_2(\Omega)}=o\left(\left(\frac{\varepsilon}{\delta}
\right)^{3/2}\right)+o(1), \qquad
\|W_\varepsilon^\delta-1\|_{L_2(\Omega)} =o(1),\\ &
\bigg|\bigg|\frac{
\partial W_\varepsilon^\delta}{\partial x_3}-\frac{\pi
c_\omega\varepsilon}{2\delta}\bigg|\bigg| _{L_2(\Gamma_1)}=
o\left(\frac{\varepsilon}{\delta}\right).
\end{align*}
\end{lemma}

Note, we consider Problem (\ref{1.4}) in a distributional sense.
Therefore, a function $u_{\varepsilon,\delta}\in H^1(\Omega;
\Gamma_2)$ is called a solution of (\ref{1.4}), if there holds the
integral identity
\begin{equation}
\int\limits_\Omega\nabla u_0\nabla
v\,dx+Q\int\limits_{\Gamma_1}u_0v\,ds= -\int\limits_\Omega
fv\,dx,\label{1.6}
\end{equation}
for any $v\in H^1 (\Omega;\Gamma_2)$, respectively.


Let $v$ be a function from $C^\infty(\Omega;\Gamma_2)$. Then
(\ref{1.2}) implies that
\begin{equation}
 \int\limits_\Omega\left(\nabla
u_{\varepsilon,\delta},\nabla\left(W_\varepsilon^ \delta
v\right)\right)\,dx+q\int\limits_{\Gamma_{\varepsilon,\delta}^S}u_
{\varepsilon,\delta}v W_\varepsilon^\delta\,ds=
-\int\limits_\Omega fvW_\varepsilon^\delta\,dx.\label{4.1}
\end{equation}

Keeping in mind the definition of ${H^1}(\Omega;S)$, the Green
formula and (\ref{4.1}) we deduce

$$-\int\limits_\Omega \Delta vW^\delta_\varepsilon
u_{\varepsilon,\delta}\,dx - \int\limits_\Omega v\Delta
W^\delta_\varepsilon u_{\varepsilon,
\delta}\,dx+I_{\varepsilon,\delta}+$$
\begin{equation}
+\int\limits_{\Gamma_{\varepsilon,\delta}^S}\left(qvW_\varepsilon^\delta+\frac{
\partial v}{\partial x_3}W_\varepsilon^\delta + \frac{ \partial
W_\varepsilon^\delta}{\partial x_3}v
\right)u_{\varepsilon,\delta}\,ds =-\int\limits_\Omega
fvW_\varepsilon^\delta\,dx, \label{4.2}
\end{equation}
where
\begin{equation}
 I_{\varepsilon,\delta}=-2
\int\limits_\Omega\left(\nabla v,\nabla
W_\varepsilon^\delta\right)u_{\varepsilon, \delta}\,dx.\label{4.3}
\end{equation}

Integrating by parts (\ref{4.3}), we obtain that

\begin{equation}
I_{\varepsilon,\delta}=2\int\limits_\Omega\left(\left(\nabla
v,\nabla u_{\varepsilon, \delta}\right)+
u_{\varepsilon,\delta}\Delta
v\right)W_\varepsilon^\delta\,dx-2\int\limits_{\Gamma_1}\left(
W_\varepsilon^\delta-1 \right)u_{\varepsilon,\delta}\frac{
\partial v}{\partial x_3}\,ds.\label{4.4}
\end{equation}

Let $\{\varepsilon_n\}$ be a sequence which tends to zero as
$n\to\infty$. Due to the embedding theorems, weak compactness of
the bounded set in $H^1(\Omega)$ and estimate (\ref{3.1}), there
exists a subsequence $\{\varepsilon'_n\}$, such that
$u_{\varepsilon,\delta}\to u_0\in H^1(\Omega;\Gamma_2)$ weakly in
$H^1(\Omega)$ and strongly in $L_2(\Omega)$ as $\varepsilon=
\varepsilon'_n\to 0$. Using Lemma~\ref{lm4.2} and passing to the
limit in (\ref{4.2}), and (\ref{4.4}) as $\varepsilon'_n\to 0$, we
obtain that

\begin{equation}
-\int\limits_\Omega \Delta
vu_0\,dx+\int\limits_{\Gamma_1}u_0\frac{
\partial v}{
\partial x_3}\, ds + Q\int\limits_{\Gamma_1}u_0v\,ds
=-\int\limits_\Omega fv\,dx. \label{4.5}
\end{equation}

Using the Green formula and (\ref{4.5}), we obtain (\ref{1.6}). By
means of the definition of ${H^1}(\Omega;S)$ the function $u_0$ is
a solution of (\ref{1.4}). On the other hand, due to the
arbitrariness of choosing of the sequence $\{\varepsilon_n\}$, we
obtain that $u_{\varepsilon,\delta}\to u_0$ as $\varepsilon\to0$
(weakly in $H^1(\Omega)$ and strongly in $L_2(\Omega)$). For
$p=0$, the strong convergence of $u_{\varepsilon,\delta}\to u_0$
in  $H^1(\Omega)$ proves similarly as in the proof of
Theorem~\ref{th1.2}. Theorem is proved.

\section{Proof of Theorem~\ref{th1.4}}\label{s5}



\begin{definition}
The function $\widetilde{u}_0\in{H^1}
(\widetilde\Omega\backslash\overline\Omega;
\Gamma\cup\widetilde\Gamma)$ is a solution of (\ref{5.3}), if it
satisfies the identity
\begin{equation}
 \int\limits_{\widetilde\Omega\backslash\overline\Omega}
 \nabla\widetilde  u_0\nabla v\,dx=
-\int\limits_{\widetilde\Omega\backslash\overline\Omega}
\widetilde fv\,dx,\label{A.1}
\end{equation}
for any $v\in{H^1} (\widetilde\Omega\backslash\overline\Omega;
\Gamma\cup\widetilde\Gamma)$.
\end{definition}

Note, that for boundary -- value problem  (\ref{5.1}) (for boundary --
value problem (\ref{6.1})), the surface
$\Gamma_{\varepsilon,\delta}^D\cup\Gamma_2$ (the surface
$\Gamma_2\cup\Gamma_{\varepsilon,\delta}^S$) is considered as
two-sides. Keeping in mind this remark, a function
$u_{\varepsilon,\delta}\in{H^1}(\widetilde\Omega
\backslash\overline
{\Gamma_{\varepsilon,\delta}^D\cup\Gamma_2};\widetilde\Gamma\cup
{\Gamma_{\varepsilon,\delta}^D\cup\Gamma_2})$, or
$u_{\varepsilon,\delta}\in{H^1}(\widetilde\Omega\backslash\overline{\left(\Gamma_2\cup\Gamma_{\varepsilon,\delta}^S
\right)};\widetilde\Gamma\cup {\Gamma_2})$ is called a solution of
(\ref{5.1}), or (\ref{6.1}), if there holds the integral identity
\begin{equation}
\int\limits_\Omega\nabla u_{\varepsilon,\delta}\nabla
v\,dx+\int\limits_{\widetilde\Omega\backslash\overline\Omega}\nabla
u_{\varepsilon,\delta}\nabla v\,dx=
-\int\limits_{\widetilde\Omega} Fv\,dx\label{A.2}
\end{equation}
for any $v\in{H^1}(\widetilde\Omega \backslash\overline
{\Gamma_{\varepsilon,\delta}^D\cup\Gamma_2};\widetilde\Gamma\cup
{\Gamma_{\varepsilon,\delta}^D\cup\Gamma_2})$, or
$v\in{H^1}(\widetilde\Omega\backslash\overline{\left(\Gamma_2\cup\Gamma_{\varepsilon,\delta}^S
\right)};\widetilde\Gamma\cup {\Gamma_2})$, respectively.

Using the integral identity~(\ref{A.2})  of Problem~(\ref{5.1})
and keeping in mind the definition of the space ${H^1}
(\widetilde\Omega;\widetilde\Gamma\cup\Gamma_{\varepsilon,
\delta}^D)\cup\Gamma_2$, we conclude that the uniform estimates

\begin{equation}
\|u_{\varepsilon,\delta}\|_{H^1(\Omega)}\le
C\|F\|_{L_2(\widetilde\Omega)}\label{5.4}
\end{equation}
and
\begin{equation}
\|u_{\varepsilon,\delta}\|_{H^1(\widetilde\Omega\backslash\Omega)}\le
C\|F\|_{L_2(\widetilde\Omega)}\label{5.5}
\end{equation}
hold true.

\begin{lemma}\label{lm5.1} For any
$v\in{H^1}(\widetilde\Omega;\widetilde\Gamma\cup
\Gamma_{\varepsilon,\delta}^D\cup\Gamma_2)$ the following
estimates:
\begin{equation}
\|v\|_{L_2(\Gamma)}\le
C\left(\frac{\delta}{\varepsilon}\right)^{1/2}\|v\|_{H^1(\Omega)}\label{5.6}
\end{equation}
and
\begin{equation}
 \|v\|_{L_2(\Gamma)}\le
C\left(\frac{\delta}{\varepsilon}\right)^{1/2}\|v\|_{H^1
(\widetilde\Omega\backslash\Omega)}\label{5.7}
\end{equation}
are valid.
\end{lemma}

The proof of this Lemma is similar to the proof of
Lemma~\ref{lm3.1}.

Let $p=\infty$, suppose also that $v$ is a function from
$C^\infty_0(\Omega)$ and $\{\varepsilon_n\}$ is a sequence which
tends to zero as $n\to\infty$. Due to the embedding theorems, weak
compactness of the bounded set in $H^1(\Omega)$ and estimates
(\ref{5.4}), (\ref{5.6}) there exists a subsequence of this
sequence such that $u_{\varepsilon,\delta}\to u_0\in {H^1}
(\Omega;\Gamma)$ at this subsequence weakly in $H^1(\Omega)$ and
strongly in $L_2(\Omega)$. Multiplying the equation (\ref{A.2}) of
Problem (\ref{5.1}) by a test--function from
${H^1}(\widetilde\Omega;\widetilde\Gamma\cup
\Gamma_{\varepsilon,\delta}^D\cup\Gamma_2)$, integrating over
$\Omega$ and passing to the limit on this subsequence we obtain
the integral identity (\ref{1.5}) of Problem (\ref{1.3}). Hence,
$u_0$ is a solution of (\ref{1.3}). On the other hand, due to the
arbitrariness of choosing of the sequence of $\{\varepsilon_n\}$,
we have that $u_{\varepsilon,\delta}\to u_0$ as $\varepsilon\to0$
weakly in $H^1(\Omega)$ and strongly in $L_2(\Omega)$. The strong
convergence of $u_{\varepsilon,\delta}\to u_0$ in  $H^1(\Omega)$
proves similarly as in the proof of Theorem~\ref{th1.2}.

Acting in the same way, we obtain the convergence of
$u_{\varepsilon,\delta}$ in the domain
$\widetilde\Omega\backslash\Omega$. Theorem is proved.

\begin{remark}\label{rm5.1} It is easy to see that the statement of Theorem holds true if
we consider instead of the fixed function $F$ the oscillating
function $F_{\varepsilon,\delta}$ such that
$F_{\varepsilon,\delta}\rightharpoonup F$ weakly in
$L_2(\widetilde{\Omega})$.
\end{remark}



\section{Proof of Theorem~\ref{th1.5}}\label{s6}

Denote by $\widehat{W}^\varepsilon(x)$ the even continuation of
the function
\begin{equation*}
1-  \chi(r\varepsilon^{-1/2}) X_0(x\varepsilon^{-1})
\end{equation*}
defined in  $\Sigma$, with respect to $x_3$. We conserve the same
notation for the $\sigma$-periodic translation  of the function
$\widehat{W}^\varepsilon(x)$ on the plane $x_3=0$.
 Taking into account the definition of $X_{0}$
one can obtain the following statement.

\begin{lemma}\label{lm6.1} Let $p=0$,
$\widehat{W}^{\varepsilon,\delta}(x)=\widehat{W}^\varepsilon\left(\frac{x}{\delta}\right)$.
Then $\widehat{W}^{\varepsilon,\delta}\in
H^1(\widetilde{\Omega};\Pi^{\delta}_{\varepsilon})$,
\begin{equation*}
\|\widehat{W}^{\varepsilon,\delta}-1\|_{H^{1}(\widetilde{\Omega})}=o(1),
\quad\varepsilon\to0.
\end{equation*}
\end{lemma}

Let $v$ be a function from $C^\infty(\Omega;\Gamma_2)$,
$\widetilde{v}$ be a function from
$C^\infty(\widetilde{\Omega}\backslash\overline{\Omega};\widetilde{\Gamma}\cup\Gamma_2)$.
Then, due to lemma~\ref{lm6.1}
\begin{equation}
\|v-\widehat{W}^{\varepsilon,\delta}v\|_{H^1(\Omega)}+\|\widetilde{v}-
\widehat{W}^{\varepsilon,\delta}\widetilde{v}\|_{H^1(\widetilde{\Omega}\backslash\overline{\Omega})}
\underset{\varepsilon\to0}\to0,\label{6.0}
\end{equation}
the continuation of $\widehat{W}^{\varepsilon,\delta}v$ into
$\widetilde{\Omega}\backslash\overline{\Omega}$ by zero belongs
$H^1(\widetilde{\Omega}\backslash\overline{\Gamma_2\cup\Gamma^S_{\varepsilon,\delta}};\widetilde{\Gamma}\cup
\Gamma_2)$, and the continuation of
$\widehat{W}^{\varepsilon,\delta}\widetilde{v}$ into $\Omega$ by
zero belongs
$H^1(\widetilde{\Omega}\backslash\overline{\Gamma_2\cup\Gamma^S_{\varepsilon,\delta}};\widetilde{\Gamma}\cup
\Gamma_2)$, too. So, (\ref{A.2}) can be written as
\begin{equation}
\int\limits_\Omega\left(\nabla
u_{\varepsilon,\delta},\nabla\left(\widehat{W}^{\varepsilon,\delta}
v\right)\right)\,dx= -\int\limits_\Omega
fv\widehat{W}^{\varepsilon,\delta}\,dx\label{6.4}
\end{equation}
or
\begin{equation}
\int\limits_{\widetilde{\Omega}\backslash\overline{\Omega}}\left(\nabla
u_{\varepsilon,\delta},\nabla\left(\widehat{W}^{\varepsilon,\delta}
\widetilde{v}\right)\right)\,dx=
-\int\limits_{\widetilde{\Omega}\backslash\overline{\Omega}}
\widetilde{f}\widetilde{v}\widehat{W}^{\varepsilon,\delta}\,dx,\label{6.41}
\end{equation}
where $f$ and $\widetilde{f}$ are the restrictions of $F$ on
$\Omega$ and $\widetilde{\Omega}\backslash\overline{\Omega}$,
respectively.

From the integral identity~(\ref{A.2}) we could deduce the uniform
boundedness of the function $u_{\varepsilon,\delta}$ in
$H^1(\widetilde{\Omega})$. Let $\{\varepsilon_n\}$ be a sequence
which tends to zero as $n\to\infty$. Due to the embedding
theorems, weak compactness of the bounded set of functions in
$H^1(\widetilde{\Omega})$, there exists a subsequence
$\{\varepsilon'_n\}$, such that $u_{\varepsilon,\delta}\to u_*$
weakly in $H^1(\Omega)$ and strongly in $L_2(\Omega)$ as
$\varepsilon'_n\to 0$, $u_*\in H^1(\Omega;\overline{\Gamma}_{2})$,
and $u_{\varepsilon,\delta}\to \widetilde{u}_*$ weakly in
$H^1(\widetilde{\Omega}\backslash\overline{\Omega})$ and strongly
in $L_2(\widetilde{\Omega}\backslash\overline{\Omega})$ as
$\varepsilon'_n\to 0$, $\widetilde{u}_*\in
H^1(\widetilde{\Omega}\backslash\overline{\Omega};\overline{\Gamma}_{2})$.

Keeping in mind (\ref{6.0}) and passing to the limit in
(\ref{6.4}), and (\ref{6.41}) as $\varepsilon'_n\to 0$, we obtain
\begin{equation}
\int\limits_\Omega\left(\nabla u_*,\nabla v\right)\,dx=
-\int\limits_\Omega fv,dx\label{6.8}
\end{equation}

and

\begin{equation}
\int\limits_{\widetilde{\Omega}\backslash\overline{\Omega}}\left(\nabla
u_*,\nabla\widetilde{v}\right)\,dx=
-\int\limits_{\widetilde{\Omega}\backslash\overline{\Omega}}
\widetilde{f}\widetilde{v}\,dx\label{6.81}
\end{equation}
respectively.  Due to the uniqueness of solutions to Problems
(\ref{6.2}) and (\ref{6.3}) we conclude that $u_*\equiv u_0$ and
$\widetilde{u}_*\equiv \widetilde{u}_0$.

On the other hand, due to the arbitrariness of choosing of the
sequence $\{\varepsilon_n\}$, we obtain that
$u_{\varepsilon,\delta}\to u_0$ as $\varepsilon\to0$ (weakly in
$H^1(\Omega)$ and strongly in $L_2(\Omega)$) and
$u_{\varepsilon,\delta}\to \widetilde{u}_0$ weakly in
$H^1(\widetilde{\Omega}\backslash\overline{\Omega})$ and strongly
in $L_2(\widetilde{\Omega}\backslash\overline{\Omega})$ as
$\varepsilon\to 0$.

The proof of the strong convergence of $u_{\varepsilon,\delta}\to
u_0$ in $H^1(\Omega)$ and $u_{\varepsilon,\delta}\to
\widetilde{u}_0$ in
$H^1(\widetilde{\Omega}\backslash\overline{\Omega})$ is similar to
that one in Theorem~\ref{th1.2}. Theorem is proved.

\begin{remark}\label{rm6.3}
It is easy to see that the statement of Theorem holds true if we
consider instead of the fixed function $F$ the oscillating
function $F_{\varepsilon,\delta}$ such that
$F_{\varepsilon,\delta}\rightharpoonup F$ weakly in
$L_2(\widetilde{\Omega})$.
\end{remark}

\section{ Construction of analytic continuations of
solutions}\label{s7}


In this section we give the construction of the solutions of the
problems (\ref{1.10})--(\ref{1.13}) and their analytic
continuations. This construction reproduces enough standard
construction \cite{bib26} given in \cite{bib28}, for the Helmholtz
resonator and in \cite{bib29}--\cite{bib31} for its
two-dimensional analogue in homogenization.

If $X$ is a notation for some Banach space (for instance,
$X=L_2$), then $ X_{loc}(D)\overset{def}{=}\{u:\, u\in X(D\cap
S(R))\quad\forall R\}$, where $S(R)$ is the open ball of radius
$R$ centered at the origin. We say that a sequence converges in
$X_{loc}(D)$, if it converges in $X(D\cap S(R))$ for all $R$. Let
$\mathcal{B}(X,Y)$ be the Banach space of bounded linear operators
mapping the Banach space $X$ into the Banach space $Y$, ${\mathcal
B}(X)\overset{def}{=}{\mathcal B}(X,X)$,
$\mathcal{B}(Y,X_{loc}(D))$ be the set of maps ${\mathcal
A}\,:Y\to X_{loc}(D)$ such that ${\mathcal A}\in{\mathcal
B}(Y,X(D\cap S(R)))$ for all $R$. We indicate by ${\mathcal
B}^h(X,Y)$ (by ${\mathcal B}^m(X,Y)$) the set of holomorphic
(meromorphic) operator-valued functions whose values belong to
${\mathcal B}(X,Y)$; ${\mathcal
B}^{h(m)}(X,X)\overset{def}{=}{\mathcal B}^{h(m)}(X)$, ${\mathcal
B}^{h(m)}(X,Y_{loc}(D))\overset{def}{=}\{\mathcal A:\, \mathcal
A\in {\mathcal B}^{h(m)}(X,Y(D\cap S(R)))\quad\forall R\}$.

Let $L$ be any number such that $\overline{\Omega}\subset S(L/3)$.
So, we can put $\widetilde{\Omega}=S(L)$. Let's  consider two
families of boundary -- value problems in bounded domains:
\begin{align}
\left\{
\begin{aligned}
&\Delta u_{\varepsilon,\delta}=\Delta w,\quad
x\in\widetilde\Omega\backslash\overline{\Gamma_{\varepsilon,\delta}^D\cup\Gamma_2},
\cr
 & u_{\varepsilon,\delta}=0,
 \quad x\in\Gamma_{\varepsilon,\delta}^D\cup\Gamma_2,\qquad u_{\varepsilon,\delta}=w,\quad x\in\widetilde
 \Gamma,
\end{aligned}\right.
\label{7.1}
\end{align}
and
\begin{align}
 \left\{
\begin{aligned}
& \Delta u_{\varepsilon,\delta}=\Delta w,\quad
x\in\widetilde\Omega\backslash\overline{\left(\Gamma_2\cup\Gamma_{\varepsilon,\delta}^S
\right)}, \cr & u_{\varepsilon,\delta}=0,\quad x\in\Gamma_2,\qquad
u_{\varepsilon,\delta}=w,\quad x\in \widetilde\Gamma, \cr &
\dfrac{\partial u_{\varepsilon,\delta}}{\partial x_3}=0, \quad
x\in\Gamma_{\varepsilon,\delta}^S,
\end{aligned}\right.\label{7.2}
\end{align}
where $w\in H^2(\widetilde{\Omega})$. Denote by
$\sigma_\varepsilon^{(1)}$ an operator whose value on $w\in
H^2(\widetilde{\Omega})$ is the solution
$u_{\varepsilon,\delta(\varepsilon)}\in
H^1(\widetilde\Omega\backslash\overline{\Gamma_{\varepsilon,\delta}^D\cup\Gamma_2})$
of the problem (\ref{7.1}) and  denote by
$\sigma_\varepsilon^{(2)}$ an operator whose value on $w\in
H^2(\widetilde{\Omega})$ is the solution
$u_{\varepsilon,\delta(\varepsilon)}\in
H^1(\widetilde\Omega\backslash\overline{\left(\Gamma_2\cup\Gamma_{\varepsilon,\delta}^S
\right)})$ of the problem (\ref{7.2}). Similarly, denote by
$\sigma_0^{(1)}$ an operator whose value on $w\in
H^2(\widetilde{\Omega})$ is the pair of the solution $u_0\in
H^1(\Omega)$ and $\widetilde{u}_0\in
H^1(\widetilde\Omega\backslash\overline{\Omega})$ of the problems
\begin{align}
\left\{
\begin{aligned}
 \Delta & u_0=\Delta w,\quad x\in\Omega, \cr & u_0=0,\quad x\in\Gamma,
\end{aligned}
\right. \label{7.3}
\end{align}
and
\begin{align}
 \left\{
\begin{aligned}
& \Delta \widetilde{u}_0=\Delta w,\quad
x\in\widetilde\Omega\backslash\overline
 {\Omega}, \cr &
 \widetilde{u}_0=0,\quad x\in\Gamma,\qquad \widetilde{u}_0=w,\quad x\in\widetilde\Gamma,
\end{aligned}\right.
\label{7.4}
\end{align} respectively,  and  denote by $\sigma_0^{(2)}$ an operator
whose value on $w\in H^2(\widetilde{\Omega})$ is the pair of the
solution $u_0\in H^1(\Omega)$ and $\widetilde{u}_0\in
H^1(\widetilde\Omega\backslash\overline{\Omega})$ of the problems
\begin{align} & \left\{
\begin{aligned}
& \Delta u_0=\Delta w,\quad x\in\Omega,\cr &
 u_0=0,\quad x\in\Gamma_2, \cr &
 \dfrac{\partial u_0}{\partial x_3}=0,\quad x\in\Gamma_1,
\end{aligned}\right.
\label{7.5}
\\
& \left\{
\begin{aligned}
& \Delta \widetilde{u}_0=\Delta w,\quad
x\in\widetilde\Omega\backslash\overline{\Omega},\cr &
 \widetilde{u}_0=0,\quad x\in\Gamma_2,\qquad \widetilde{u}_0=w,\quad x\in\widetilde \Gamma, \cr &
 \dfrac{\partial \widetilde{u}_0}{\partial x_3}=0,\quad
 x\in\Gamma_1,
\end{aligned}\right.
\label{7.6}
\end{align}
respectively. Set
$$
(\Delta+k^2)^{-1}g\overset{def}{=}-\frac{1}{4\pi}\int\limits_{S(L)}\frac{\hbox{\rm
e}^{\hbox{\rm i}k|x-y|}}{|x-y|}\,g(y) \,dy,\qquad x\in{\mathbb
R}^3,
$$
Hereinafter we use the same notation for  a function from
$L_2(S(L))$ and its continuation by zero outside $S(L)$ meaning
the latter as a function from $L_2(\mathbb{R}^3)$. Next, let $p_L$
be the operator of restriction to $S(L)$,
\begin{align*}
A^{(j)}_\mu(k)\overset{def}{=}& \left(1+
\chi\left(\frac{r}{L}\right)\left(\sigma^{(j)}_\mu
p_L-1\right)\right)
(\Delta+k^2)^{-1},\\
T^{(j)}_\mu(k)g\overset{def}{=} & \Bigg(\left(\Delta+k^2\right)
\left(\chi\left(\frac{r}{ L}\right)\right) \left(\left(1
-\sigma^{(j)}_\mu\,p_L\right)\left(\Delta+k^2\right)^{-1}\right)
\\
& + 2\sum_{i=1}^2 \frac{\partial}{\partial
x_i}\left(\chi\left(\frac{r}{L}\right)\right)
\frac{\partial}{\partial x_i}\left(\left(1
-\sigma^{(j)}_\mu\,p_L\right)
\left(\Delta+k^2\right)^{-1}\right)\Bigg)g,
\end{align*}

From the definitions of $T^{(j)}_\mu(k)$ it follows that for $g\in
L_2(S(L))$, the function $T^{(j)}_\mu(k)g\in L_2({\mathbb R}^3)$
and $\mathrm{supp}\, T^{(j)}_\mu(k)g\subset\overline{S(L)}$. For
this reason, the maps $T^{(j)}_\mu(k)$ and
$B^{(j)}_\mu(k)=I-T^{(j)}_\mu(k)$ can be considered as operators
from $L_2(S(L))$ into $L_2(S(L))$. Under this interpretation, from
the definitions of $A^{(j)}_\mu(k)$ and $T^{(j)}_\mu(k)$ the
following statements hold.

\begin{lemma}\label{lm7.1} For $k\in{\mathbb
C}$,

a) mappings  $A^{(1)}_\mu(k)\in {\mathcal
B}^h\left(L_2(S(L)),{H^1_{loc}}({\mathbb R}^{3})\right)$,
\begin{equation*}
\begin{aligned}
A^{(2)}_\varepsilon(k)&\in {\mathcal
B}^h\left(L_2(S(L)),{H^1_{loc}}({\mathbb
R}^{3}\backslash\overline{\Gamma_{\varepsilon,\delta(\varepsilon)}^S})\right),\\
A^{(2)}_0(k)&\in {\mathcal
B}^h\left(L_2(S(L)),{H^1_{loc}}({\mathbb
R}^{3}\backslash\overline{\Gamma_1})\right),\\
T^{(j)}_\mu(k)&\in {\mathcal B}^h\left(L_2(S(L))\right),
\end{aligned}
\end{equation*}
and, for any fixed $k$ and $\mu$, $T^{(j)}_\mu(k)$ is a compact
operator in $L_2(S(L))$;

b) the function
$u_{\varepsilon,\delta(\varepsilon)}=A^{(1)}_\varepsilon(k)g$
satisfies
 (\ref{1.10}) for
$F=(I-T^{(1)}_\varepsilon(k))g$, where $I$ is the identity
mapping, the function
$u_{\varepsilon,\delta(\varepsilon)}=A^{(2)}_\varepsilon(k)g$
satisfies
 (\ref{1.11}) for
$F=(I-T^{(2)}_\varepsilon(k))g$, the restriction $u_0$ of
$A^{(1)}_0(k)g$ to $\Omega$ satisfies (\ref{1.33}), where $f$ is
the restriction of $(I-T^{(1)}_0(k))g$ to  $\Omega$,
 the restriction $\widetilde{u}_0$ of
$A^{(1)}_0(k)g$ to $\mathbb{R}^3\backslash\overline{\Omega}$
satisfies (\ref{1.12}), where $\widetilde{f}$ is the restriction
of $(I-T^{(1)}_0(k))g$ to $S(L)\backslash\overline{\Omega}$,
 the restriction $u_0$ of
$A^{(2)}_0(k)g$ to $\Omega$ satisfies (\ref{6.22}), where $f$ is
the restriction of $(I-T^{(2)}_0(k))g$ to  $\Omega$,
 the restriction $\widetilde{u}_0$ of
$A^{(2)}_0(k)g$ to $\mathbb{R}^3\backslash\overline{\Omega}$
satisfies (\ref{1.13}), where $\widetilde{f}$ is the restriction
of $(I-T^{(2)}_0(k))g$ to $S(L)\backslash\overline{\Omega}$, and,
for $\mathrm{Im}\, k\ge0$, the functions
$u_{\varepsilon,\delta(\varepsilon)}$ and $\widetilde{u}_0$ also
satisfies the radiation conditions (\ref{rad1}), (\ref{rad2}).
\end{lemma}

The square root of the eigenvalue is called the eigenfrequency of
the boundary -- value problem. Denote by $\Sigma^{(1)}$ and
$\Sigma^{(2)}$ the sets of eigenfrequencies of boundary -- value
problems (\ref{1.33}) and (\ref{6.22}), respectively.

\begin{proposition}\label{pr7.1}  If $\hbox{\rm Im\,} k\ge0$
then the solution of the perturbed problem {\rm (\ref{1.10})
((\ref{1.11})), (\ref{rad1})} and the limit external problem {\rm
(\ref{1.12}) ((\ref{1.13})), (\ref{rad2})} are unique. If
$k\notin\Sigma^{(1)}$ ($k\notin\Sigma^{(2)}$) then the solution of
the limit internal problem {\rm (\ref{1.33}) ((\ref{6.22}))} is
unique.
\end{proposition}

The proof of this statement is wellknown (for the
three-dimensional external Neumann problem outside nonclosed
surfaces see, for instance in \cite{bib28}).

\begin{lemma}\label{lm7.2} If $g\not=0$ and $g\in L_2(S(L))$, then $A_\mu^{(j)}(k)g\not=0$.
\end{lemma}

The proof of this statement is completely identical to the proof
of Lemma 2.2 in \cite{bib28} and Lemma 3.3 in \cite{bib31}.

From Proposition~\ref{pr7.1},  Lemma~\ref{lm7.2} and the
compactness of the operator $T^{(j)}_\mu(k)$ we deduce the
following Lemma.

\begin{lemma}\label{lm7.3} If $\hbox{\rm Im\,} k>0$ or $k>0$, and
$k\notin\Sigma^{(j)}$,  then there exists the operator
$\left(B_\mu^{(j)}\right)^{-1}(k)\in{\mathcal{B}}(L_2(S(L)))$.
\end{lemma}

Later we will use the following statement from \cite{bib26}

\begin{proposition}\label{pr7.2}
Suppose that $D$ is a connected domain of the complex plane,
$T(k)$ ($k\in D$) is a holomorphic family of compact operators in
a Banach space $\mathcal{X}$ and there exists a point $k_*\in D$,
such that $(I-T(k_*))^{-1}\in \mathcal{B}(\mathcal{X})$. Then
$(I-T(k))^{-1}$ is a meromorphic function in $D$ with values in
$\mathcal{B}(\mathcal{X})$.
\end{proposition}

From Proposition~\ref{pr7.2} and Lemma~\ref{lm7.3} the following
Lemma follows:

\begin{lemma}\label{lm7.4} $\left(B_\mu^{(j)}\right)^{-1}(k)\in\mathcal{B}^{m}(L_2(S(L)))$ in $\mathbb{C}$.
\end{lemma}

Denote \qquad $ {\mathcal
A}_\mu^{(j)}(k)\overset{def}{=}A_\mu^{(j)}(k)
\left(B_\mu^{(j)}\right)^{-1}(k) $

\begin{theorem}\label{th7.1} For $k\in{\mathbb
C}$,

a) mappings  $\mathcal{A}^{(1)}_\mu(k)\in {\mathcal
B}^m\left(L_2(S(L)),{H^1_{loc}}({\mathbb R}^{3})\right)$,
\begin{equation*}
\begin{aligned}
\mathcal{A}^{(2)}_\varepsilon(k)&\in {\mathcal
B}^m\left(L_2(S(L)),{H^1_{loc}}({\mathbb
R}^{3}\backslash\overline{\Gamma_{\varepsilon,\delta(\varepsilon)}^S})\right),\\
\mathcal{A}^{(2)}_0(k)&\in {\mathcal
B}^m\left(L_2(S(L)),{H^1_{loc}}({\mathbb
R}^{3}\backslash\overline{\Gamma_1})\right);
\end{aligned}
\end{equation*}

b) the function
$u_{\varepsilon,\delta(\varepsilon)}=\mathcal{A}^{(1)}_\varepsilon(k)F$
satisfies
 (\ref{1.10}), the function
$u_{\varepsilon,\delta(\varepsilon)}=\mathcal{A}^{(2)}_\varepsilon(k)F$
satisfies
 (\ref{1.11}), the restriction $u_0$ ($\widetilde{u}_0$) of
$\mathcal{A}^{(1)}_0(k)F$ to $\Omega$ (to
$\mathbb{R}^3\backslash\overline{\Omega}$) satisfies (\ref{1.33})
((\ref{1.12})), where $f$ ($\widetilde{f}$) is the restriction of
$F$ to  $\Omega$ (to $S(L)\backslash\overline{\Omega}$),
 the restriction $u_0$ ($\widetilde{u}_0$) of
$\mathcal{A}^{(2)}_0(k)F$ to $\Omega$ (to
$\mathbb{R}^3\backslash\overline{\Omega}$) satisfies (\ref{6.22})
((\ref{1.13})), where $f$ ($\widetilde{f}$) is the restriction of
$F$ to  $\Omega$ (to $S(L)\backslash\overline{\Omega}$),  and, for
$\mathrm{Im}\, k\ge0$, the functions
$u_{\varepsilon,\delta(\varepsilon)}$ and $\widetilde{u}_0$ also
satisfies the radiation conditions (\ref{rad1}), (\ref{rad2});

c) if $\hbox{supp\,} F\subset S(T)$, then the function
$\mathcal{A}^{(j)}_\varepsilon(k)F$ does not depend on $L\ge T$;
$j=1,2;$

d) the set of poles of the operators
$\mathcal{A}^{(j)}_\varepsilon(k)$ and
$\left(B^{(j)}_\mu(k)\right)^{-1}$ coincide for fixed $j$;
$j=1,2;$

e) the set of poles of the operator
$\mathcal{A}^{(j)}_\varepsilon(k)$ does not depend on $L$;
$j=1,2.$

\end{theorem}

\noindent\textbf{Proof.} The statements a) and b) follow from
Lemmas~\ref{7.1}, \ref{7.4}. The statement c) follows from
Proposition~\ref{7.1} and the uniqueness of the analytic
continuation. The statement d) follows from Lemma~\ref{7.2} and
the definition of the operators
$\mathcal{A}^{(j)}_\varepsilon(k)$; $j=1,2$. Let us prove the
statement e) for $j=1$. Denote by $\mathcal
A^{(1)}_{\varepsilon,t}$ the operator $\mathcal
{A}^{(1)}_{\varepsilon}$ defined for $L=t$. Suppose $a>b$. It is
obvious that the set of poles of $\mathcal A^{(1)}_{\varepsilon,
b}$ is a subset of the set of poles of $\mathcal
A^{(1)}_{\varepsilon, a}$. Now we show the inverse inclusion.
Suppose that $\mathrm{supp} F\subset S(a)$ and assume that
$$
W=(1-\chi(r b^{-1}))(\Delta+k^2)^{-1} F.
$$
Since the support $\mathrm{supp} (F-(\Delta+k^2)W)\subset
\overline{S(b)}$, then due to b) the solution of the perturbed
problem (\ref{1.10}), (\ref{rad1}) can be defined by one of the
following formulae
$$
u_{\varepsilon,\delta}=\mathcal A^{(1)}_{\varepsilon,a} (k)
F,\qquad u_{\varepsilon,\delta}=\mathcal A^{(1)}_{\varepsilon,a}
(k)(F-(\Delta+k^2)W)+W.
$$
Since $W$ is holomorphic the set of poles of $\mathcal
A^{(1)}_{\varepsilon,a}$ is a subset of the set of poles of
$\mathcal A^{(1)}_{\varepsilon,b}.$ The case $j=2$ can be proved
in analogues way. Theorem is proved.

\section{ Proof of Theorems~\ref{th1.6} and \ref{th1.7}}\label{s8}

By the definition of operators $\sigma_\delta^{(j)}$, Statements
 of Theorems~\ref{th1.4} , \ref{th1.5} and Remarks~\ref{rm5.1} and
\ref{rm6.3} we deduce the following Lemma:

\begin{lemma}\label{lm8.1} If $p=\infty$, then
$$\|\sigma_\varepsilon^{(1)}-\sigma_0^{(1)}\|_{
\mathcal{B}(W_2^2(S(L)),W_2^1(S(L)))}\underset{\varepsilon\to0}\to0.$$
If $p=0$,then
$$\|\sigma_\varepsilon^{(2)}-\sigma_0^{(2)}\|_{\mathcal{B}(W_2^2(S(L)),W_2^1(S(L)\backslash\overline{\Omega}\cup\Omega))}
\underset{\varepsilon\to0}\to0.$$
\end{lemma}

\begin{definition}
Suppose that $\mathcal{D}_\mu$ is a family of operators acting
from the Banach space $\mathcal{X}$ in $L_{2,loc}(\mathbb{R}^3)$
(in $H^1_{loc}(\mathbb{R}^3)$,
$H^1_{loc}(\mathbb{R}^3\backslash\overline{\Omega}$).

We say that $\mathcal{D}_\varepsilon\underset{\varepsilon\to0}\to
\mathcal{D}_0$ strongly (weakly, by the norm) in
$\mathcal{B}(\mathcal{X},L_{2,loc}(\mathbb{R}^3))$ (in
${\mathcal{B}}(\mathcal{X},H_{loc}^1(\mathbb{R}^3))$,
$\mathcal{B}(\mathcal{X},H_{loc}^1(\mathbb{R}^3\backslash\overline{\Omega})$),
if $\mathcal{D}_\varepsilon\underset{\varepsilon\to0}\to
\mathcal{D}_0$ strongly (weakly, by the norm) in
$\mathcal{B}(\mathcal{X},L_{2}(S(R))$ (in
$\mathcal{B}(\mathcal{X},H^1(S(R)))$,
$\mathcal{B}(\mathcal{X},H^1(S(R)\backslash\overline{\Omega}))$)
for any $R>0$.
\end{definition}

The next Lemma follows from the definitions of operators
$A_\mu^{(j)}(k)$, $T_\mu^{(j)}(k)$, and Lemma~\ref{lm8.1}:

\begin{lemma}\label{lm8.2}  Assume that $K$ is an arbitrary compact set in $\mathbb{C}$. Then

a) if $p=\infty$, then
$A_\varepsilon^{(1)}(k)\underset{\varepsilon\to0}\to A_0^{(1)}(k)$
by the norm $\mathcal{B}(L_2(S(L)),H_{2}^1(\mathbb{R}^3))$ and
$T_\varepsilon^{(1)}(k)\underset{\varepsilon\to0}\to T_0^{(1)}(k)$
by the norm $\mathcal{B}(L_2(S(L)))$ uniformly on $k\in K$;

b) if $p=0$, then
$A_\varepsilon^{(2)}(k)\underset{\varepsilon\to0}\to A_0^{(2)}(k)$
by the norm
$\mathcal{B}(L_2(S(L)),H_{loc}^1(\mathbb{R}^3\backslash\overline{\Omega}\cup\Omega))$
and $T_\varepsilon^{(2)}(k)\underset{\varepsilon\to0}\to
T_0^{(2)}(k)$ by the norm $\mathcal{B}(L_2(S(L)))$ uniformly on
$k\in K$.
\end{lemma}

Later on we shall use the following statement from \cite{bib26}:

\begin{proposition}\label{pr8.1} Suppose that $D$ is a connected domain in the complex
plane, $T(k,\mu)$ is a family of compact operators in Banach space
$\mathcal{X}$, defined for $k\in D$ and $\mu\in[0,\mu_0]$, such
that it is holomorphic on  $k$ for each $\mu$ and continuous on
$D\times[0,\mu_0]$ by the norm $\mathcal{B}(\mathcal{X})$.
Furthermore assume that there exists the point $k_0\in D$, such
that $(I-T(k_0,\mu))^{-1}\in \mathcal{B}(\mathcal{X})$ for any
$\mu\in(0,\mu_0)$. Then

a) $(I-T(k,\mu))^{-1}$ (for any $\mu$) is a meromorphic function
in $D$ with values in $\mathcal{B}(\mathcal{X})$;

b) if $k_*$ is not a pole $(I-T(k,\mu_*))^{-1}$, then the
operator-values function $(I-T(k,\mu))^{-1}$ is continuous by the
norm in a neighborhood of $(k_*,\mu_*)$;

c) the poles $(I-T(k,\mu))^{-1}$ depend on $\mu$ continuously.
\end{proposition}

Denote by $\Sigma^{(j)}_{\mu}$ the set of the poles of the
operator $\mathcal{A}^{(j)}_{\mu}$. By Lemmas~\ref{lm7.4} and
\ref{lm8.2} the family $T_\mu^{(j)}(k)$ satisfies the conditions
of Proposition \ref{pr8.1}. Then the following Lemma holds true:

\begin{lemma}\label{lm8.3} a) Assume that  $p=\infty$. If
$K$ is an arbitrary compact set in $\mathbb{C}$, such that
$K\cap\Sigma_0^{(1)}=\emptyset$, then
$\left(B_\varepsilon^{(1)}(k)\right)^{-1}\underset{\varepsilon\to0}
\to \left(B_0^{(1)}(k)\right)^{-1}$ by the norm
$\mathcal{B}(L_2(S(L)))$ uniformly on $k\in K$. If
$\tau_0\in\Sigma_0^{(1)}$, then there exists the pole
$\tau_\varepsilon\in\Sigma_\varepsilon^{(1)}$, converging to
$\tau_0$ as $\varepsilon\to0$.

b) Assume that $p=0$. If $K$ is an arbitrary compact set in
$\mathbb{C}$, such that $K\cap\Sigma_0^{(2)}=\emptyset$, then
$\left(B_\varepsilon^{(2)}(k)\right)^{-1}\underset{\varepsilon\to0}
\to \left(B_0^{(2)}(k)\right)^{-1}$ by the norm
$\mathcal{B}(L_2(S(L)))$ uniformly on $k\in K$. If
$\tau_0\in\Sigma_0^{(2)}$, then there exists the pole
$\tau_\varepsilon\in\Sigma_\varepsilon^{(2)}$, converging to
$\tau_0$ as $\varepsilon\to0$.

\end{lemma}

Finally from Lemmas~\ref{lm8.2} and \ref{lm8.3} we deduce.

\begin{theorem}\label{th8.1}
a) Suppose that $p=\infty$. If $K$ is an arbitrary compact set in
$\mathbb{C}$, such that $K\cap\Sigma_0^{(1)}=\emptyset$, then
$\mathcal{A}_\varepsilon^{(1)}(k)\underset{\varepsilon\to0}\to
\mathcal{A}_0^{(1)}(k)$ by the norm of the space
$\mathcal{B}(L_2(S(L)),H^1_{loc}(\mathbb{R}^3))$ uniformly on
$k\in K$. If $\tau_0\in\Sigma_0^{(1)}$, then there exists a pole
$\tau_\varepsilon\in\Sigma_\varepsilon^{(1)}$, converging to
$\tau_0$ as $\varepsilon\to0$.

b) Suppose that $p=0$. If $K$ is an arbitrary compact set in
$\mathbb{C}$, such that $K\cap\Sigma_0^{(2)}=\emptyset$, then
$\mathcal{A}_\varepsilon^{(2)}(k)\underset{\varepsilon\to0}\to
\mathcal{A}_0^{(2)}(k)$ by the norm
$\mathcal{B}(L_2(S(L)),H^1_{loc}(\mathbb
R^{3}\backslash\overline{\Gamma_{1}}))$ uniformly on $k\in K$. If
$\tau_0\in\Sigma_0^{(2)}$, then there exists a pole
$\tau_\varepsilon\in\Sigma_\varepsilon^{(2)}$, converging to
$\tau_0$ as $\varepsilon\to0$.
\end{theorem}

Theorems~\ref{th1.6} and \ref{th1.7} are the implication of
Theorems~\ref{th8.1} and \ref{th7.1}.

\section*{Acknowledgments.}

The paper was completed in Narvik University College (HiN) in the
Northern Norway during the cold January 2003. The authors express
thanks to Narvik University College for the wonderful conditions
to work and for the support.



\end{document}